\documentclass[a4paper,twocolumn,showpacs,superscriptaddress,floatfix]{revtex4-1}

\usepackage{graphicx}
\usepackage{epsf}
\usepackage{epsfig}
\usepackage{amssymb,amsmath}
\usepackage[usenames]{color}
\usepackage{amssymb}
\usepackage{times}
\usepackage{tabularx}
\usepackage[table]{xcolor}

\newcommand{\be}{\begin{equation}}
\newcommand{\ee}{\end{equation}}
\newcommand{\bea}{\begin{eqnarray}}
\newcommand{\eea}{\end{eqnarray}}

\font\tenscr=rsfs10 scaled1100
\font\sevenscr=rsfs7 
\font\fivescr=rsfs5 
\skewchar\tenscr='177
\skewchar\sevenscr='177
\skewchar\fivescr='177
\newfam\scrfam
\textfont\scrfam=\tenscr
\scriptfont\scrfam=\sevenscr
\scriptscriptfont\scrfam=\fivescr

\def\scri{{\fam\scrfam I}}

\begin{document}

\title{Gravitational wave signatures of black hole quasi-normal mode instability }

\author{Jos\'e Luis Jaramillo}
\affiliation{Institut de Math\'ematiques de Bourgogne (IMB), UMR 5584, CNRS, Universit\'e de
Bourgogne Franche-Comt\'e, F-21000 Dijon, France}
\author{Rodrigo Panosso Macedo}
\affiliation{School of Mathematical Sciences, Queen Mary, University of
  London, \\ Mile End Road, London E1 4NS, United Kingdom}
\affiliation{CENTRA, Departamento de F\'{\i}sica, Instituto Superior T\'ecnico -- IST, Universidade de Lisboa -- UL, Avenida Rovisco Pais 1, 1049 Lisboa, Portugal}
\author{Lamis Al Sheikh}
\affiliation{Institut de Math\'ematiques de Bourgogne (IMB), UMR 5584, CNRS, Universit\'e de
Bourgogne Franche-Comt\'e, F-21000 Dijon, France}

\begin{abstract}
Black hole (BH) spectroscopy has emerged as a powerful approach to extract spacetime information from gravitational wave (GW) observed signals. Yet, 
quasinormal mode (QNM) spectral instability under high wave-number perturbations has been recently shown to be a common classical general relativistic phenomenon~\cite{Jaramillo:2020tuu}. This requires to assess its impact on the BH QNM spectrum, in particular on BH QNM overtone frequencies. We conclude: i) perturbed BH QNM overtones are
indeed potentially observable in the GW waveform,
providing information on small-scale environment BH physics, and ii) their detection poses a challenging data analysis problem of singular interest for LISA astrophysics.  We adopt a two-fold approach, combining theoretical results from scattering theory with a fine-tuned data analysis on a highly-accurate numerical GW ringdown signal. The former introduces a set of effective parameters (partially lying on a BH Weyl law) to characterise QNM instability physics. The latter provides a proof-of-principle demonstrating that the QNM spectral instability is indeed accessible in the time-domain GW waveform, though certainly requiring large signal-to-noise ratios. Particular attention is devoted  to discuss the patterns of isospectrality loss under QNM instability, since the disentanglement between axial and polar GW parities may already occur within the near-future detection range. 

\end{abstract}


\maketitle
\noindent
 {\bf Introduction:} Are {\em all} black-hole vibrational modes observables in gravitational-wave astronomy? What astrophysical/fundamental physics information do they {\em really} convey?

 Gravitational waves (GW) from binary systems are systematically observed by current GW antennae~\cite{Abbott:2020niy}. The late-time radiation of newly formed  black holes (BHs) is typically characterised by an exponentially damped, oscillating signal. The so-called quasi-normal modes (QNMs) encode the decaying scales and oscillating frequencies. An indispensable tool in astrophysics, fundamental gravitational physics, and mathematical relativity~\cite{Chandrasekhar:579245,Kokkotas:1999bd,Nollert:1999ji,Berti:2009kk,Konoplya:2011qq}, the QNMs provide structural information about the BH's background. The future generation of ground- and space-based detectors shall provide data sufficiently accurate to measure several QNMs~\cite{Berti:2005ys,Dreyer:2003bv,Baibhav_2018,Ota:2019bzl,Isi:2019aib,Giesler:2019uxc,Isi:2020tac,Cabero:2019zyt,Maggio:2020jml,Ota:2021ypb}, allowing one to address fundamental questions in physics~\cite{Barack:2018yly,Barausse:2020rsu}. 
 
 Small environmental perturbations are not expected to radically disrupt the underlying BH spacetime, given the confidence in BH dynamical stability. Yet, instabilities seem intrinsic to the theory at the  spectral level~\cite{Jaramillo:2020tuu,Berti:2021ijo,Gasperin:2021kfv}. Fluctuations may alter significantly the QNM spectrum itself, with stronger effects in the high overtones~\cite{Jaramillo:2020tuu}. Since recent GW events hints towards the detectably of more than one mode~\cite{Giesler:2019uxc,Isi:2019aib,Isi:2020tac,Capano:2021etf}, addressing our opening questions is paramount for the correct interpretation of future GW observations. 

\begin{figure}[b!]
\centering
\includegraphics[width=7.4cm]{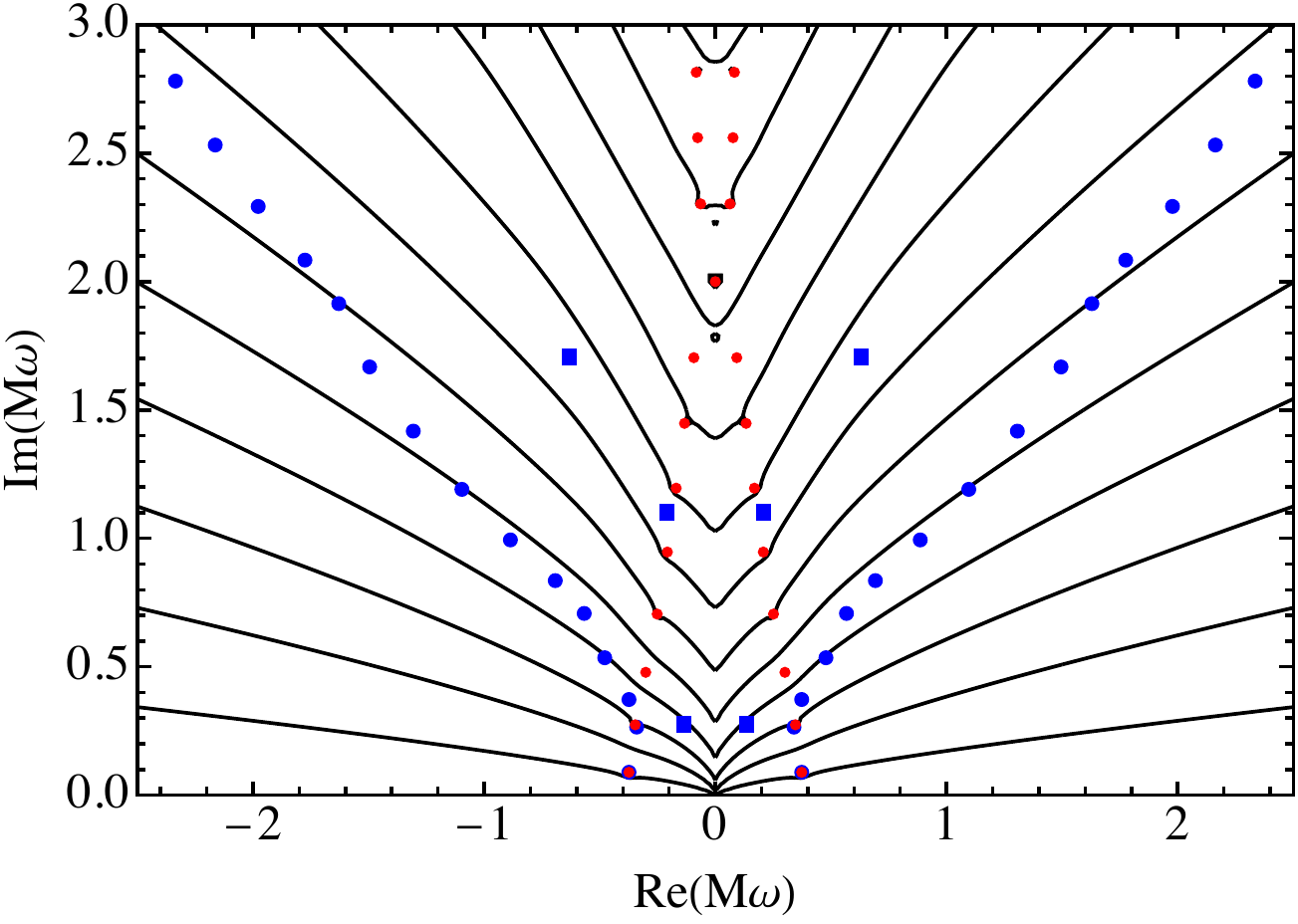}
\caption{
  The pseudospectra contourlines (black) identify regions in the complex plane where the eigenvalues may move under perturbations of the underlying non-self-adjoint operator. In BH spacetimes, they spread open, signalising spectral instabilities. Under perturbations $\epsilon \, \delta V_k$ characterised by small amplitude $\epsilon \ll1$ and large wave-number $k\gg1$, the QNM overtones differ significantly from their original values (in red circles). One observes a branch sharing the tendency set by the pseudospectra~\cite{Jaramillo:2020tuu} (blue circles) and new internal modes (blue squares) with decay scale similar to some QNM in the branch, but lower oscillatory frequency. 
Here, $\epsilon=10^{-3}$ and $k=10$.
 }
\label{fig:reviewPRX}
\end{figure}

 BH perturbation theory is described via a nonconservative system with energy leaking inside the BH and propagating out to the wave zone. Equations are described by non-self-adjoint operators, in a common framework across classical and quantum systems~\cite{Ashida:2020dkc}. The notion of pseudospectra, recently introduced into gravity~\cite{Jaramillo:2020tuu}, allows to identify spectral instabilities in nonconservative systems~\cite{TreTreRed93,trefethen2005spectra,Sjostrand2019,Ashida:2020dkc}. As a topographical map, the pseudospectra contour level with value $\epsilon$ delimits the region in the complex plane where QNMs can migrate when the system undergoes perturbations of this order. The spectra is stable if such $\epsilon$-contour lines lie within a distance $\epsilon$ from the original spectra. Spectral instability follows when the contour lines extend into a large region in the complex plane. Remarkably, the latter arises in BH physics~\cite{Jaramillo:2020tuu,Berti:2021ijo,Destounis:2021lum,Gasperin:2021kfv} (see fig.~\ref{fig:reviewPRX}).

 \begin{figure*}[t!]
\centering
\includegraphics[height=6.2cm]{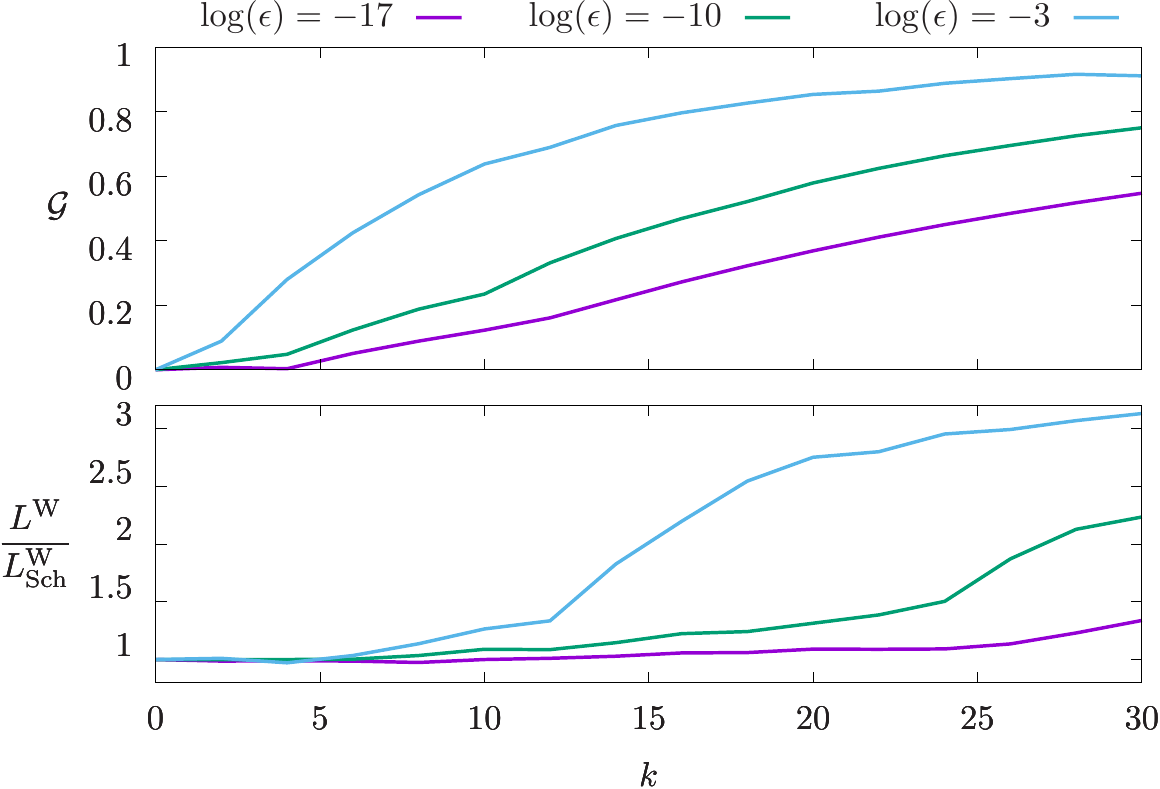}
\includegraphics[height=6.2cm]{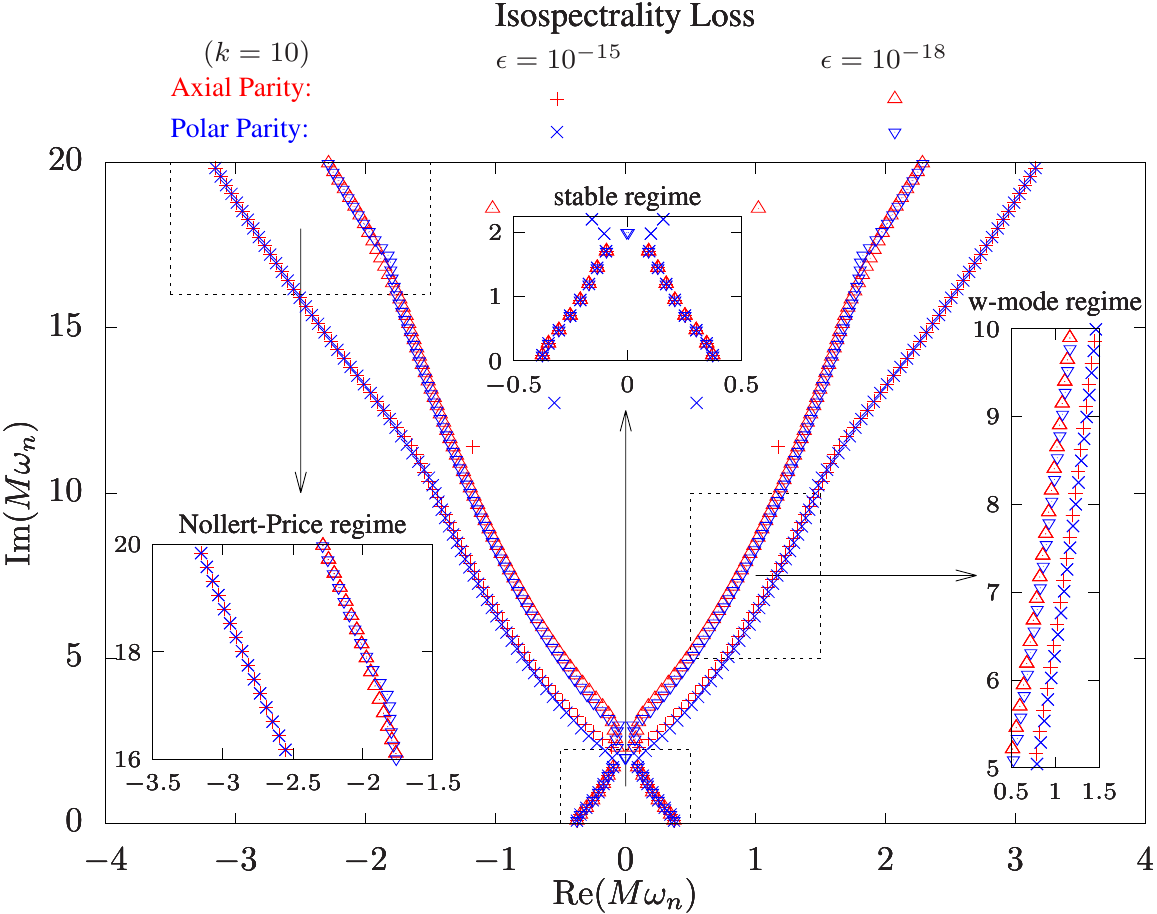}
\caption{
 {\em Left Panel:} Effective measures accounting for QNMs' distribution. Perturbed potential amplitudes $\epsilon=10^{-17}, 10^{-10}, 10^{-3}$. Branch opening is assessed by ${\cal G}=\lim_{n\to\infty}\omega^{\rm R}_n/|\omega_n|$ (top). Schwarzschild QNMs have ${\cal G}=0$, whereas eq.~\eqref{e:Regge_branches} yields ${\cal G}=1$. The tendency ${\cal G}\rightarrow1$ as $k\rightarrow \infty$ indicates QNMs migrating to $\epsilon-$pseudospectra log-lines in the large wave-number limit. The Weyl's law length $L^{\rm W}$ (bottom) follows from counting the number of QNMs within a delimited region in the complex plane. The transition $L^{\mathrm{W}}/L^{\mathrm{W}}_{\mathrm{Sch}} =1$ to $O(3)$ follows from QNMs internal to the branches becoming densely populated. Thus ${\cal G}$ and $L^{\rm W}$ are complementary measures assessing generic features in QNM instabilities (c.f.~fig.~\ref{fig:reviewPRX}). {\em Left Panel:} Regimes of axial (red) - polar (blue) isospectrality loss ($k=10$ and $\epsilon= 10^{-15}$, $10^{-18}$). In the stable regime (top inset), lower QNMs are not affected by instability, thus axial and polar QNMs differs with order $\epsilon$. The ``$w$-mode" regime (right inset) strongly distinguishes axial and polar QNMs into an an alternating pattern. In the ``Nollert-Price" regime (left inset) axial and polar QNMs differs again just within order $\epsilon$, despite the wider branch opening. Transition between the regimes occurs close to internal modes.
  }
\label{fig:stable_isospec_loss}
\end{figure*}

 {\bf QNM instability:} To trigger the instabilities, ref.~\cite{Jaramillo:2020tuu} introduced an {\it ad hoc} modification $\epsilon \,\delta V_k$  ($\epsilon \ll 1$)  into the potential governing the dynamics of GWs on a spherically symmetric BH spacetime. When having a sinusoidal profile in the radial direction,  $\epsilon \, \delta V_k$ mimics a Fourier mode from a realistic potential and it captures the contribution of small and large scale perturbations via a wave-number $k$. Fig.~\ref{fig:reviewPRX} reproduces the overtones instability for $k\gg1$~\cite{Jaramillo:2020tuu}. We stress the appearance of: (i) branches opening similarly to the pseudospectra lines (blue circles), dubbed ``Nollert-Price" branches~\cite{Nollert:1996rf,Nollert:1998ys} by ref.~\cite{Jaramillo:2020tuu}; (ii) modes (blue squares) inside the region fixed by the  ``Nollert-Price" branches, named here ``internal modes".

Specific values for the perturbed QNMs depend on the particular model for the environmental effects or modifications in the gravity theory. Yet, the opening pattern observed in fig.~\ref{fig:reviewPRX} is rather generic, which raises the need for a research program aiming at understanding the GW observational implications of such QNM instabilities. The challenge lies on several fronts. 

On the theoretical side, apart from modelling the specificities of the local  environmental astrophysics, or extending gravity beyond General Relativity (e.g.~\cite{Yunes:2013dva,Perkins:2020tra}), GW astronomy shall profit from formal results in the theory of scattering resonances~\cite{Regge58,LaxPhi71,LaxPhi89,Vainb73,Sjoes90,Marti02,SjoZwo07,Zworski99,zworski2017mathematical,dyatlov2019mathematical,BinZwo}. The perturbed QNM patterns reflect features agnostic to the model under consideration. From the data analysis side, one may need enhanced detection pipelines so that the features displayed in fig.~\ref{fig:reviewPRX} are not overseen, if present.

{\bf Effective parameters:} We initiate this research line on the theoretical side by adapting to gravity further results from the theory of scattering resonances. Consistently with scattering theory~\cite{Regge58,LaxPhi71,LaxPhi89,Vainb73,Sjoes90,Marti02,SjoZwo07}, our numerical analysis demonstrate the logarithmic asymptotics of pseudospectra contour lines (cf. also ~\cite{Jaramillo:2020tuu,Destounis:2021lum}). Moreover, 
``Nollert-Price" QNM branches open up in the complex plane in qualitatively similar patterns~\cite{Jaramillo:2020tuu}. Which information lies in the asymptotics? Currently, the question is not an observational one, as it would require an ideal instrument able to detect QNM overtones $\omega_n$ with
$n\gg1$. It is rather of structural nature: asymptotics do offer a guideline to identify the relevant patterns in the phenomenon~\cite{Batte01}.

If the dynamics of the physical scenario is dictated by potentials with discontinuities at some $p$th-derivative (i.e., of class $C^p$), then the spectra asymptotics must  reach {\em exactly} the logarithmic boundaries of the pseudospectra. The real $\omega^{\rm R}_n$ and imaginary $\omega^{\rm I}_n$ parts follow ``Regge QNM branches''~\cite{Regge57,Zwors87}
\bea
\label{e:Regge_branches}
\!\!\!\!\!\omega^{\rm R}_n \sim \pm\frac{\pi}{L^{\rm R}}\left( n + \tilde \gamma \right), \,\,
\omega^{\rm I}_n \sim \frac{1}{L^{\rm R}} \bigg[\gamma\ln \left( \left|\omega_n^{\rm R}\right| + \gamma'\right) - \ln S\bigg],
\eea
for $n\gg 1$. Reverberations within chambers with a length scale $L^{\rm R}$ is the mechanism behind the opening of the spectra into such log-branches~\cite{Regge57,Zwors87,Berry82}. These are modulated by `regularity' $\gamma,  \tilde \gamma, \gamma'$ and `strength' $S$ parameters. This behaviour is found, for instance, in BH-like potentials~\cite{Nollert:1996rf,Nollert:1998ys,Qian:2020cnz,Liu:2021aqh}, and in the $w$-modes of (a class of) neutron stars~\cite{ZhaWuLeu11,Berry82,BerMou72}. 

Detecting QNMs obeying eq.~\eqref{e:Regge_branches} would be a strong indication of an underlying low regularity ($C^p$-) potential. This feature of the $n\gg1$ QNM-pattern, suggests the introduction of a set of effective parameters: a reverberation length scale $L^{\mathrm{R}} := \pi/|\Delta\omega^R_{n}|$; a `small-scale' structure $\gamma:= L^{\mathrm{R}} \Delta\omega^{\rm I}_n/\Delta \ln \omega^{\rm R}_n$; and a perturbation strength $\ln S_n := \gamma\ln(\omega^{\rm R}_n) -  L^{\mathrm{R}} \omega^{\rm I}_n$~\footnote{The definitions directly recover the parameters from the spiked truncated dipole potential~\cite{Nollert:1998ys}, namely $L\sim x_\delta- x_0$ (length of `cavity') and $S\sim V_\delta$ (potential amplitude). The polytropic neutron stars~\cite{ZhaWuLeu11} have $L \sim r^*$ (star's radius), $\gamma \sim {\cal N}$ (polytropic index) and $S\sim$ ``discontinuity jump of the potential''.}.

Rigourous results for the spectra distribution (within the $\epsilon-$pseudospectra region) for smooth potentials are less sharp, but {\em QNMs must always lay above the logarithmic curves}. We conjecture that the QNMs reach the log-curves in the large $k$ wave-number limit. Supporting this statement, we introduce ${\cal G}_n = \omega^{\rm R}_n/|\omega_n|$ to measure the opening. This is a different representation of the so-called quality factor $Q_n$ (e.g.~\cite{LalYanVyn17}; cf.~\cite{Pook-Kolb:2020jlr} in BH QNM physics). Schwarzschild QNMs' asymptotics~\cite{Nollert:1993zz} gives $ {\cal G}:=\lim_{n\to\infty} {\cal G}_n = 0$, whereas eq.~\eqref{e:Regge_branches} yields  ${\cal G} = 1$. Fig.~\ref{fig:stable_isospec_loss}'s upper-left panel shows the monotonic increase of ${\cal G}\in[0,1]$
  (for several $\epsilon$'s) with  $k$. The tendency ${\cal G} \rightarrow 1$ as $k\to\infty$ is a strong indication that the pseudospectra's log-boundaries are attained in the large wave-number limit.
  
 {\bf Weyl law:}  
 The Weyl law is a spectral concept common across physical theories
 ~\cite{BaltesHilf,Berger03,Arendt2009}, but scantly explored in (scattering) GW physics~\cite{Moss02}.
 Rigorous results for a QNM Weyl's law are 
 available for scenarios modelled by potentials with compact support or of class $C^p$~\cite{Zwors87,Froes97,Simon00,Sjost14}. Let $N(\omega)$ be the number of QNMs in the radius $|\omega_n|<\omega$ ($\omega\in \mathbb{R}$). For one-dimensional potentials, the
 QNM Weyl's law states $N(\omega) \sim 2 (L^{\mathrm{W}}/\pi)\omega$, with $L^{\mathrm{W}}$  a length scale of the potential. 

 Typical potentials in BH perturbation theory 
do not satisfy the hypotheses of theorems' above,
but we do observe that BH QNMs indeed follow a Weyl law~\cite{JarMacRaf21}. The Schwarzschild's QNM asymptotics~\cite{Nollert:1993zz} for an angular mode $\ell$ yields  $N_\ell(\omega) \!\!=\!\!8M\omega$, i.e. a scale $L^{\mathrm{W}}_{\mathrm{Sch}} \! = \! 4\pi M$.
This connects with the exploration of BH horizon area quantisation and BH thermodynamics based on QNM asymptotics ~\cite{Moss02,Motl:2002hd,Motl:2003cd,Maggi08}, with a link to Hawking temperature via $2L^{\mathrm{W}}_{\mathrm{Sch}}=(T_{\rm Hawking})^{-1}$. Finally, summing $N_{\ell}(\omega)$ over $(\ell, m)$ yields $N(\omega)\sim \omega^3$, in accordance with Weyl's law in $d=3$ spatial dimensions~\cite{Sjost14}, probing the effective spacetime dimension by counting QNMs.

Weyl Law's remains valid for perturbed BH potentials and $L^{\mathrm{W}}$ is always robustly defined (fig.~\ref{fig:stable_isospec_loss}'s lower-left panel). The changes in $L^{\mathrm{W}}$ are not related to the branch opening. Indeed, we observe $|\Delta \omega_n|$ constant along them. Rather, the apparent `phase transition' with 'order parameter' $L^{\mathrm{W}}/L^{\mathrm{W}}_{\mathrm{Sch}}$ shifting from $1$ to $O(3)$ results from an increase of internal QNMs. 

Hence, the measure of ${\cal G}$ and the Weyl length scale $L^{\mathrm{W}}$ are complementary to each other, as they assess precisely the two novel aspects in QNM instability~\footnote{$L^{\mathrm{W}}$ and ${\cal G}$ are always defined, in contrast with  Regge's length $L^{\mathrm{R}}$ in eq. (\ref{e:Regge_branches}).
If the latter is also defined, it holds $L^{\mathrm{R}} = L^{\mathrm{W}}{\cal G}$.}: ${\cal G}$ accounts for the ``Nollert-Price" branches opening in the complex plane, whereas $L^{\mathrm{W}}$ measures the presence of internal modes.
 
{\bf Isospectrality loss:}  
Another outcome of the QNM overtones instability is the distinction between axial and polar GW parities. While both QNM spectra coincide for the Schwarzschild BH, parity disentanglement is a natural consequence when the system is slightly perturbed~\cite{Jaramillo:2020tuu}. We observe the existence of three regimes of the isospectrality loss (fig.~\ref{fig:stable_isospec_loss}):

(i) {\em Stable region:} Relatively low wave-number $k$ and small perturbed amplitude $\epsilon$ do not trigger the instability in the first few QNM overtones. The perturbed polar/axial  QNMs stay at a distance $\epsilon$ from their original values. Isospectrality loss is then of the same order $\epsilon$ as the perturbation, i.e. $|\omega^{\mathrm{axial}}_n - \omega^{\mathrm{polar}}_n| \sim C_n(k) \epsilon$. The function $C_n(k) \sim C_n(k + k_n)^{\alpha_n}$ has model-dependent constants $C_n$, $k_n$ and $\alpha_n$. As $\epsilon$ or $k$ increase, the stable behaviour is observed by less and less overtones, eventually reducing only to the fundamental mode. Near-future observations shall measure both parities in the fundamental QNM, which may discriminate from other physical mechanisms behind the isospectrality loss (e.g.~\cite{Cardoso:2019mqo,Maggio:2020jml}). 

(ii) {\em  Alternating axial/polar ``$w$-modes'':}  Moving to higher overtones, parities drastically separate
when QNMs instability first occurs.
QNMs of different parity place themselves in an alternating pattern along the branch, as neutron star ``$w$-modes'' do \cite{Kokkotas:1999bd,ZhaWuLeu11}.  Isospectrality loss is most accessible here, with BHs as compact star mimickers. We observe $\omega^{\rm R}_n \! \sim \! \ln (\omega^{\rm I}_n)$, $\omega^{\rm I}_n\sim n$ (cf. the contrast with 
eq.~\eqref{e:Regge_branches}). As $\epsilon$ or $k$ increases, this regime descends in $\mathbb{C}$  towards the first overtones, eventually overcoming the previous stable region.

(iii) {\em Nollert-Price regime:} In this third regime, the QNMs migrate further away from unperturbed ones. 
We observe the branches obeying $\omega_n^{\rm I} \sim \omega_n^{\rm R}\sim n$, the QNMs instability (assessed by the opening of the branch) is stronger than for alternating ``$w$-modes''. Yet, the isospectrality loss is once again linear in $\epsilon$, as in the stable regime (i). The mechanism behind this result is  unclear. This regime is the dominant one when $\epsilon$ or $k$ is sufficiently large, as it overtakes both (i) and (ii).

Interestingly, the transition between the three sectors seems to occur precisely upon appearance of an internal mode. New regimes in far asymptotic regions are not excluded, but their numerical study is challenging. We observe the internal QNMs to be very parity-sensitive, with values around the first overtone already for moderate wave-number perturbations.

Since QNM instabilities are not restricted to the asymptotic behaviour of QNMs overtones, novel features might already be within near-future detection range. The next section initiates the discussion from a simple data analysis perspective by measuring the perturbed QNMs within a numerical GW time signal. Throughout the section, {\em unbarred} quantities and results displayed in red will denote dynamics under the {\em unperturbed} potential. Equivalently, {\em barred} symbols (with results in blue) refer to dynamics under the {\em perturbed} potential. 

\begin{center}
  \begin{table*}[t!]
  \caption{QNMs for unperturbed and perturbed Schwarzschild potentials via Prony's method. Crosses are QNMs not identified.}
\begin{tabular}{|c|c|c|c|c|}
 \hline
 \multicolumn{5}{|c|}{\cellcolor{red!15} Unperturbed Potential $(\epsilon = 0,\quad k =0)$}  \\
  \hline
QNMs & $ M \omega_0$  & $M \omega_1$ & $M \omega_2$ & $M \omega_3$ \\
\hline
   Theory & $ \pm 0.37367168 - 0.08896231\, i$ & $\pm 0.3467110 - 0.2739149\,i$ & $\pm 0.3010534 - 0.4782770 \,i$ & $ \pm 0.2515049 - 0.7051482 \,i$ \\
\hline
   Prony's Fit & $\pm 0.37367169 - 0.08896232\, i$ & $\pm 0.34670 - 0.27392\, i$ & $\pm 0.302 - 0.48 \, i$ & $\times \times \times$ \\
\hline
\hline
 \multicolumn{5}{|c|}{\cellcolor{blue!15} Perturbed Potential $(\epsilon = 10^{-3},\quad k =10)$}  \\
 \hline
QNMs & $ M \overline\omega_0$  & $M \overline\omega_1$ & $M \overline\omega_2$ & $M \overline\omega_3$ \\
\hline
   Theory & $\pm 0.37364032 - 0.08898850 \,i$ & $\pm 0.3401722 - 0.2648723 \,i$ & $\pm 0.1367705 - 0.2761794  \,i$ & $ \pm 0.3735536 - 0.3723973 \,i$ \\
\hline
   Prony's Fit & $\pm 0.37364030 - 0.08898850 \, i$ & $\pm 0.342 - 0.266\, i$ & $\times \times \times$ & $\pm 0.37 - 0.4 \, i $ \\
\hline
\end{tabular}
\label{table:QNM_Prony}
\end{table*}
\end{center}

\vspace{-1.cm}
{\bf Data analysis:} Because the discussed QNMs instabilities are restricted to the overtones~\cite{Jaramillo:2020tuu}, one does not expect to see their effect in a time signal by a mere ``naked-eye" study. Indeed, the (stable) fundamental mode typically dominates the dynamics. Also, the example in fig.~\ref{fig:reviewPRX} is rather conservative in the sense that the instability is triggered only for overtones with $n\geq 2$. But interestingly, the perturbed spectra show an internal-mode $\overline \omega_2$ near the first overtone: ${\rm Im}(\overline\omega_1)\sim {\rm Im}(\overline\omega_2)$. As a proof-of-principle for eventual realistic detections (with large signal-to-noise ratio), we simulate here an ideal ringdown signal. The goal is to assess the detectability of the two classes of perturbed modes (Noller-Price and internal modes).

\begin{figure}[b!]
\centering
\includegraphics[width=7.4cm]{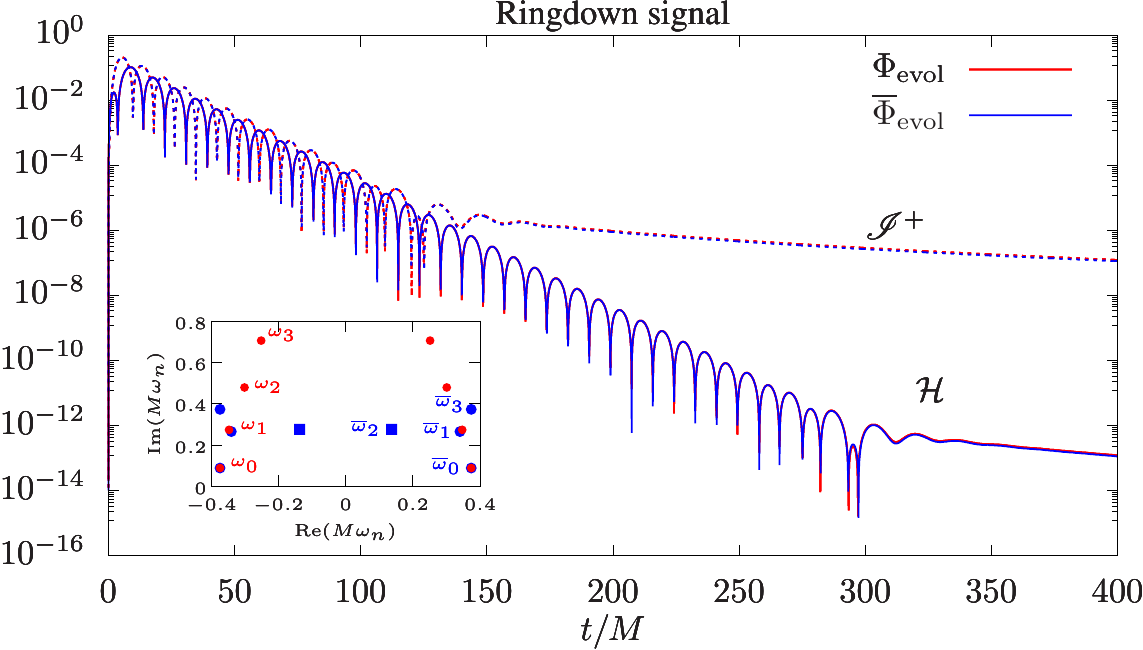}
\caption{
Numerical GW signal under unperturbed $\Phi_{\rm evol}$ (red) and perturbed $\overline \Phi_{\rm evol}$ (blue) potentials in the Schwarzschild spacetime. QNM overtone instability displayed in Fig.~\ref{fig:reviewPRX} --- here, the inset shows the first four QNMs with their labelling --- is not apparent due to the stability of fundamental mode. Prony's method infers the first overtones and detect the underlying differences in the spectra (Table~\ref{table:QNM_Prony}). 
 }
\label{fig:Schwarzschild}
\end{figure}

For this purpose, we solve the usual unperturbed Regge-Wheeler wave equation~\cite{Regge57}, as well as its perturbed version~\cite{Jaramillo:2020tuu} with $\epsilon=10^{-3}$ and $k=10$. The solutions are obtained with the highly-accurate code from ref.~\cite{Macedo:2014bfa}, which ensures the numerical noise to be at machine roundoff error.  The overall qualitative behaviour is independent of the initial data (ID). We use the so-called polynomial ID~\cite{Ansorg:2016ztf} as they ensure a QNM spectral decomposition for all times~\cite{Ansorg:2016ztf}. Fig.~\ref{fig:Schwarzschild} shows the GW time evolution at the BH horizon ${\cal H}$ and at future null infinity $\scri^+$  on the Schwarzschild background under the unperturbed (red) and perturbed (blue) Regge-Wheeler potential.

As expected, the bare signals look indistinguishable and we resort to the Prony's method to measure the QNMs~\cite{Berti:2007dg}. Table~\ref{table:QNM_Prony} compares the theoretical QNM values against those from the Prony's fitting~\footnote{For optimisation, we select the signal at ${\cal H}$ as the ring-down lasts longer than at $\scri^+$. Prony's method is applied for $\tau\in[6,50]$, with $p=8$, $L=40$, $N=401$ (ref.~\cite{Berti:2007dg}'s notation).}. We can infer three modes in both cases, but significant digits are lost on the overtones. Though the accuracy suffices to distinguish unperturbed from perturbed spectra, the method is oblivious to the internal mode.

We stress that the time signal $\overline\Phi_{\rm evol} (t)$ follows from the direct integration of the underlying wave equation with a given ID. Alternative and independently, a spectral analysis of the QNMs yields
$
\overline\Phi^{N}_{\rm spec}(t) := \sum_{n=0}^{N} \overline{\cal A}_n e^{i\overline\omega_n t },
$
with the factors $\overline{\cal A}_n$ encoding
how a given ID excites each individual QNM. 

Such ``frequency domain'' approach permits to tighten our assessment of the spectral instability within the dynamical waveform, in particular by focusing on the internal mode $\overline{\omega}_2$. Namely, we use a semi-analytical tool~\cite{Ansorg:2016ztf,Ammon:2016fru,PanossoMacedo:2018hab} to measure $\overline{\cal A}_n$. A close  look at the excitation factors reveals that the internal mode $\overline \omega_2$ is very mildly, but unmistakably, excited (see Supplemental Material): perturbed QNMs
are therefore indeed present in the perturbed GW signal. With the employed ID we get $\overline{\cal A}_2\!\sim\! 10^{-3}$, whereas  $\overline{\cal A}_0\!\sim \!\overline{\cal A}_1\!\sim\! \overline{\cal A}_3\!\sim\! 10^{-1}$. The fainter signal explains why Prony's method bypasses this mode, while its background noise spoils $\overline \omega_1$'s and $\overline \omega_3$'s accuracy. An important open question is whether more realistic ID would excite the internal modes more effectively.

{\bf Discussion:} The QNM overtone instabilities described in ref.~\cite{Jaramillo:2020tuu}
are present in GW waveforms, directly
impacting the future of BH spectroscopy. Assessing whether the instabilities are purely theoretical predictions or if realistic scenarios may trigger them is a pressing issue for the correct interpretation of future high accuracy GW observations. In this work we have: i) demonstrated that BH QNM overtone instabilities are not an artifact of the frequency-domain analysis, but they are actually present in the time-domain waveform, and ii) initiated a systematic multidisciplinary effort aiming at characterising the QNM instability signatures in GW signals.

With results from the theory of scattering resonances adapted to GW physics, we introduced new observables obtained from the QNMs' asymptotic behaviour. 
At this stage, their interest is not in the direct detection of large overtones, which is non-realistic within the near-future technology for GW astronomy. Rather, this is a theoretical contribution of fundamental nature since the effective asymptotic behaviour captures the constitutive features of the QNM instability phenomenon. In particular, such observables open an avenue to probe small-scale physics of the BH and its environment.
Complementary knowledge follows from introducing the Weyl Law in the context of GW physics, which gives hints into further classical and quantum aspects of BH physics.

Targeting near-future observations, fluctuations around the BH shall disentangle the GW spectra for axial and polar parities. For the fundamental mode the deviation is of the same order of the small perturbation. However, the first few overtones may show a significant contribution if the so-called internal modes are in the detection range. Indeed, by simulating a highly accurate GW ringdown signal, and exploiting fine-tuned fitting algorithms to measure the QNMs and to access their individual contribution into the evolution, we confirm the QNMs instabilities already in the first few overtones. 

In this work, the environmental fluctuations
  were modelled by adding an ad-hoc sinusoidal perturbation to the system (cf. ref.~\cite{Jaramillo:2020tuu}). The generality of the results follows because such approach captures the contribution from a given Fourier mode in a more realistic analysis. The time evolutions employed the particular
ID from ref.~\cite{Ansorg:2016ztf} but, since the ring-down is oblivious to the ID choice, we ensure the validity of our conclusions. Crucially, there remain open questions on whether and how more realistic scenarios trigger the instability and with which intensity each individual perturbed QNM is excited.

We stress the timely necessity for liaising the theoretical results on the fundamental aspects of BH perturbation theory with the current efforts to set goals and detection strategies for future GW missions. Detecting QNM overtones in a noisy signal already imposes a challenging data analysis task when a deterministic underlying spectrum is a priori available, as illustrated by the QNM analyses in refs.~\cite{Giesler:2019uxc,Isi:2019aib,Isi:2020tac,Capano:2021etf}. The theoretical prediction of BH QNM instability adds another layer of obstacles, since the perturbed QNM overtone  specific values will generically incorporate a stochastic component from (random) small-scale perturbations and only general patterns shall be available. This strongly indicates that only detections with very high signal-to-noise ratios will offer eligible candidates for disentagling BH overtone instabilities. In particular, this may define a challenging but tantalizing case for LISA science, requiring the development of specific data analysis tools able to cope with a more intricate parameter degeneracy.


\bigskip

\noindent{\em Acknowledgments.} We thank W. Barbe, E. Berti, N. Besset, O. Birnholtz, V. Cardoso, T. Daud\'e, K. Destounis,
E. Gasperin, B. Krishnan, O. Meneses Rojas, B. Raffaelli, O. Reula and J. Sj\"ostrand.

  This work was supported by the French ``Investissements d'Avenir'' program through
project ISITE-BFC (ANR-15-IDEX-03), the ANR ``Quantum Fields interacting with Geometry'' (QFG) project (ANR-20-CE40-0018-02), 
the EIPHI Graduate School (ANR-17-EURE-0002), 
  the Spanish FIS2017-86497-C2-1 project (with FEDER contribution),  
  the European Research Council Grant 
  ERC-2014-StG 639022-NewNGR ``New frontiers in numerical general relativity" and the European Commission
  Marie Sklodowska-Curie grant No 843152 (Horizon 2020 programme). The project used
  Queen Mary's Apocrita HPC facility, supported by QMUL Research-IT, and CCuB computational resources
  (universit\'e de Bourgogne).

\appendix
\section*{Supplemental Material}

In this supplement material, we provide support to some of the most critical physical statements in the main text.

\section{Presence of perturbed QNM overtones in the time signal}
A fundamental result in the article is the confirmation of the presence of unstable 
QNM overtones in the (perturbed) ring-down time signal. We explicitly demonstrate this key point here.
The top panel of fig.~\ref{fig:Schwarzschild} reproduces fig.~3 from the main text. As discussed there, it shows the time evolution of GWs on the Schwarzschild background measured at the BH horizon ${\cal H}$ and at future null infinity $\scri^+$ (i.e., the mathematical formalized notion of an infinitely far wave zone). Please recall from the main text that  {\em unbarred} quantities (with results displayed in red) will denote dynamics under the {\em unperturbed}~\cite{Regge57} potential, whereas, {\em barred} symbols (with results in blue) refer to dynamics under the {\em perturbed} version employed in ref.~\cite{Jaramillo:2020tuu}.

To unmistakably unveil the QNM spectral instability within the time signal, we employ the robust semi-analytical tool developed in ref.~\cite{Ansorg:2016ztf,Ammon:2016fru,PanossoMacedo:2018hab}, which is capable of identifying and filtering the contribution of each individual QNM. More specifically, we recall that the integration of the wave equation in the time domain yields a signal $\Phi_{\rm evol} (t)$ displaying two dynamical regimes: i) the QNM ring-down with an exponentially damped, oscillatory decay, and ii) a late-time  power-law decay  tail. Alternatively to such time domain approach, on can also study the underlying equation in the frequency domain. In this second approach, the QNM ring-down and tail-decay are associated, respectively, to the discrete spectrum $\{\omega_n; \phi_n(\sigma)\}$ and the continuous spectrum $\{ \omega = i y, y\geq 0; \phi(\sigma; y)\}$ of the underlying operator generating the dynamics~\cite{Ansorg:2016ztf,PanossoMacedo:2018hab} (here $\sigma$ is the relevant ---compactified--- spatial variable in the eigenfunctions $\phi$, whereas $n$ and
$y$ are discrete and continuous labels respectively parametrizing the discrete and continuous spectra). The focus here is on the discrete part of the spectrum, responsible for the ringdown decay, which is the dynamical regime currently accounted for in black-hole spectroscopy. In particular, one considers the signal ``proxy'' resulting from the spectral analysis as the (finite) superposition
\be
\label{eq:spectral_proxy}
\Phi^{N}_{\rm spec}(t) := \sum_{n=0}^{N} {\cal A}_n e^{i\omega_n t } \ .
\ee
Recall that the coefficients ${\cal A}_n$ assess how a given Initial Data (ID) excites each individual QNM in the time signal. Refs.~\cite{Ansorg:2016ztf,Ammon:2016fru,PanossoMacedo:2018hab} comprehensively discuss the algorithms to read ${\cal A}_n$ directly from the ID. Such procedure is utterly independent from the construction of $\Phi_{\rm evol} (t)$. Consistency between time and frequency domain approaches is assessed by monitoring the rests
\be
\label{eq:Filter_TimeEvol}
{\cal F}^{N}(t) = \Phi_{\rm evol}(t) - \Phi^{N}_{\rm spec}(t).
\ee

\begin{figure}[!h]
\centering
\includegraphics[width=7.4cm]{03a_Schwarzschild_TimeEvol.pdf}
\includegraphics[width=7.4cm]{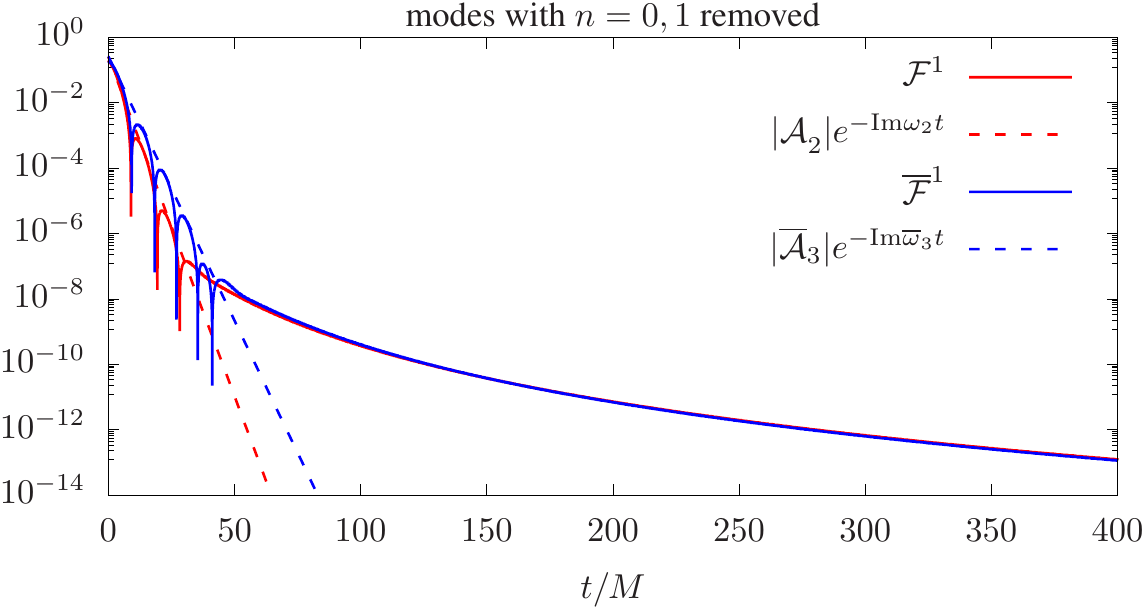}
\includegraphics[width=7.4cm]{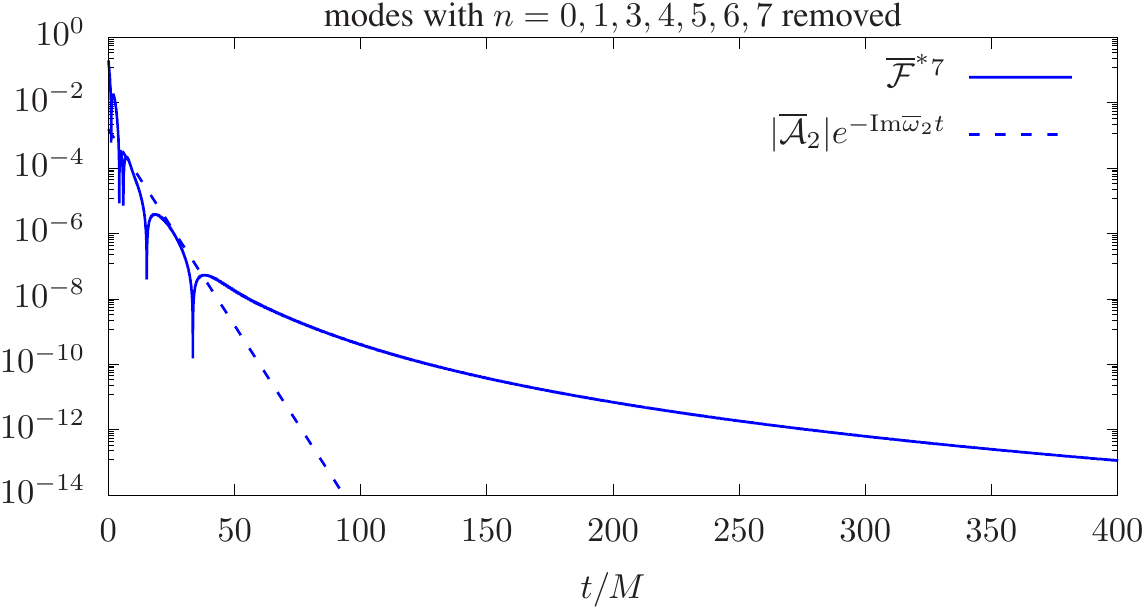}
\caption{
  {\em Top panel:} Numerical ringdown signals  $\Phi_{\rm evol}$ (red) and $\overline \Phi_{\rm evol}$ (blue) under, respectively,  unperturbed and perturbed potentials in the Schwarzschild spacetime. fig.~1 in the main text displays both underlying corresponding QNM spectra. Spectral instability is not apparent due to the stability of the dominant fundamental mode. The inset provides a zoom into the region around the fundamental QNMs and first overtones, introducing their labeling. {\em Middle panel:} Filtered ringdown signals without contribution from modes with $n=0$ and $n=1$. While ${\cal F}^1$ decays according to the corresponding unperturbed second overtone $\omega_2$, the prescribed Initial Data is not efficient to excite the internal mode $\overline \omega_2$ of the perturbed potential, and consequently $\overline {\cal F}^1$ is dominated by $\overline \omega_3$. {\em Bottom panel:} Filtered perturbed ringdown signal, with modes from $n=0$ to $n=7$ removed, except  $\overline \omega_2$. Despite the small excitation coefficient $\overline {\cal A}_2$, we indeed observe a contribution from $\overline \omega_2$ to the ringdown signal.
 }
\label{fig:Schwarzschild}
\end{figure}

By systematically filtering out the slowest decaying QNM and the first $N$ overtones from $\Phi_{\rm evol} (t)$, the filtered signal ${\cal F}^{N}(t)$ in eq.~\eqref{eq:Filter_TimeEvol} accesses higher overtones in the original full signal. Indeed, fig.~\ref{fig:Schwarzschild}'s middle panel shows the unperturbed and perturbed filtered signals without the contributions from the fundamental QNM and the first overtone, where one directly distinguishes ---in stark contrast with the full signal in the top panel--- the different dynamics for unperturbed
and perturbed filtered evolutions, respectively ${\cal F}^1$ and $\overline {\cal F}^1$. As expected, the second overtone ${\omega_2}$ dominates ${\cal F}^1$'s dynamics. However, the perturbed signal $\overline {\cal F}^1$ crucially decays exactly as predicted by $\overline \omega_3$, indeed a perturbed QNM overtone: this demonstrates the presence of the perturbed  $\overline \omega_3$ overtone in the full perturbed time signal $\overline\Phi_{\rm evol}(t)$. Note that the actually identified overtone is $\overline \omega_3$, instead of the internal mode $\overline \omega_2$. This is precisely accounted for by regarding the excitation coefficients in eq. (\ref{eq:spectral_proxy}): 
the signal from $\overline \omega_3$ is stronger than the one from $\overline \omega_2$, namely $\overline {\cal A}_3 \sim 10^{-1}$ whereas
$\overline {\cal A}_2 \sim 10^{-3}$.

But, indeed, the internal mode $\overline \omega_2$ is also present in the full perturbed signal $\overline\Phi_{\rm evol}(t)$.
To explicitly observe it, we slightly modify the theoretical filtering technique: a new filtered signal $\overline {\cal F}^*{}^N$ is built with the mode $n=2$ skipped from the sum in eq.~\eqref{eq:Filter_TimeEvol}. Fig.~\ref{fig:Schwarzschild}'s bottom panel shows $\overline {\cal F}^*{}^7$, where all modes with $n=0, \ldots, 7$ are removed, but for $n=2$. Thus, the fainter contribution from the internal mode $\overline \omega_2$ is unmistakably detected from its spectrally predicted decaying slope.

\section{Isospectrality loss: stable region}
\begin{figure}[b!]
\centering
\includegraphics[width=8cm]{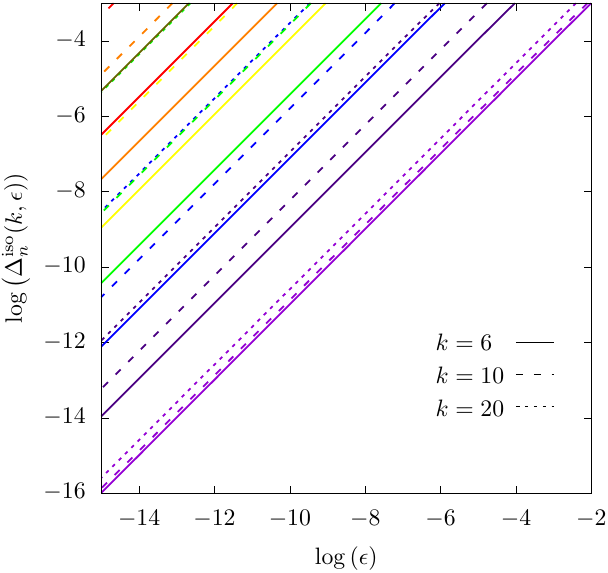}
\caption{Isospectrality loss in the stable region. Each color refers to a particular $n$ in 
  QNMs $\omega_n$: violet corresponds to the fundamental mode ($n=0$), whereas overtones ($n\geq 1$)
  correspond to the lines above. The unity slope in the log-log profile (note the same
  scale in the abscissa and the ordinate) demonstrates the strict linearity in $\epsilon$.
  The dependence of the proportionality constant $C_n(k)$ in the wave number $k$ --- cf. eq. (\ref{e:linear_iso})
  --- is illustrated by using sinusoidal perturbations with $k= 6,10,20$.
  The observed pattern corresponds to a power-law of $C_n(k) \sim C_n(k + k_n)^{\alpha_n}$, with $\alpha_n$ increasing
  with $n$.  
 }
\label{fig:Iso_loss_linear}
\end{figure}

Upon a perturbation of the order $\epsilon$, QNM isospectrality is lost for all QNMs.
But different qualitative behaviours are observed in distinct parts of the QNM spectrum.
Specifically, in the region we have referred to as ``stable'', perturbed polar and axial
QNMs stay at a distance of order $\epsilon$  from the non-perturbed QNMs.
In particular, the difference between perturbed polar and axial frequencies is also
of order $\epsilon$.
Defining
\bea
\Delta_n^{\mathrm{iso}}(k,\epsilon):=|\omega^{\mathrm{a}}_n(k,\epsilon) - \omega^{\mathrm{p}}_n(k,\epsilon)| \ , 
\eea
where $\omega^{\mathrm{a}}_n(k,\epsilon)$ and $\omega^{\mathrm{p}}_n(k,\epsilon)$ are, respectively,
the axial and polar QNM frequencies under a sinusoidal perturbation of size $\epsilon$ and
wave-number $k$, it is observed
\bea
\label{e:linear_iso}
\Delta_n^{\mathrm{iso}}(k,\epsilon) \sim C_n(k) \epsilon ,
\eea
with $C_n(k)$  presenting a power-law dependence in the wave number $k$ (namely,
$C_n(k)\sim C_n(k + k_n)^{\alpha_n}$, with constants $C_n$, $k_n$ and $\alpha_n$).
The critical point for claiming a stable behaviour is the strict linear dependence in the perturbation
size $\epsilon$.
Figure \ref{fig:Iso_loss_linear} demonstatres bluntly such behaviour. Observational access to
the differences $\Delta_n^{\mathrm{iso}}(k,\epsilon)$ would therefore directly probe the (energy)
size of the underlying perturbations.

\section{Logarithmic asymptotics of pseudospectra boundaries}
\begin{figure}[h!]
\centering
\includegraphics[width=9cm]{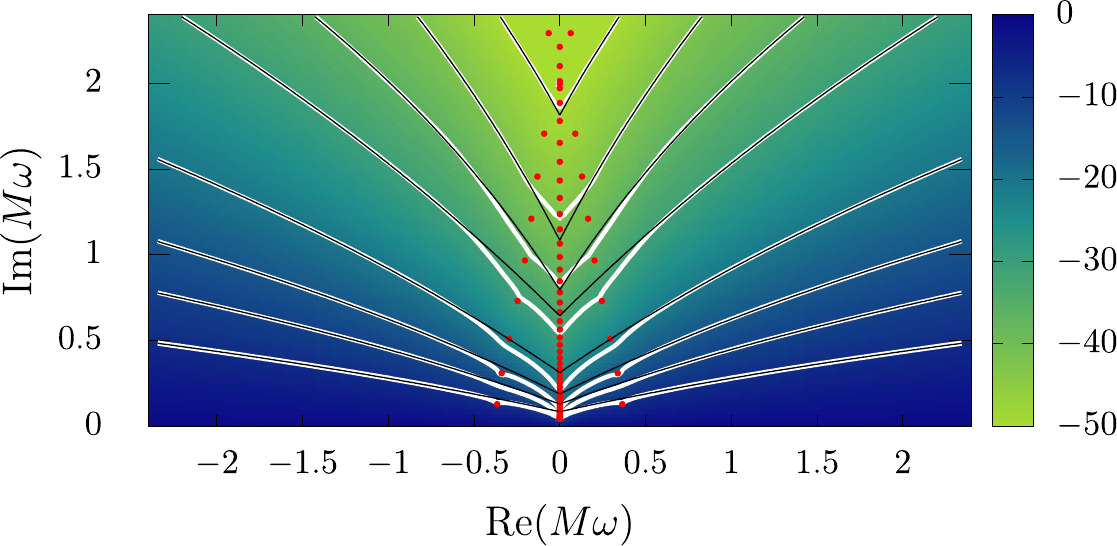}
\caption{Pseudospectrum of the Schwarzschild BH.
  The white lines correspond to
  pseudospectra boundaries, namely the level-set contour lines of the pseudospectrum
  that mark QNM-free regions and are proxies of perturbed BH QNMS (cf. ref. \cite{Jaramillo:2020tuu}).
  The thin black lines provide logarithmic fittings of such pseudospectra boundaries,
  according to eq.~\ref{e:log-asymps}. Quite remarkably, the logarithmic behaviour of pseudospectra
  boundaries extend to the region close to non-perturbed BH QNMs (red circles), therefore providing
  a parametrized pattern to model perturbed QNM overtone frequencies in data analysis.
 }
\label{fig:log-asymptotics}
\end{figure}

Data analysis strategies aiming at efficiently extracting QNM overtones from GW observational data
will possibly lean on some ``a priori'' input knowledge about the expected resonant frequencies. One of the
most striking consequences of the high-frequency overtone instability is the opening of
BH QNM branches from the asymptotically vertical ones of non-perturbed BHs to the wide-open
(Nollert-Price) branches of perturbed BHs. Pseudospectra boundaries provide ``proxies'' for
such perturbed branches (cf. fig. 1 in the main text, and \cite{Jaramillo:2020tuu}),
and present a logarithmic asymptotics for large $n$ 
\bea
\label{e:log-asymps}
\omega^{\rm I}_n \sim C_1 + C_2 \ln (\omega^{\rm R}_n+ C_3) \ .
\eea
Beyond confirming such asymptotics, the present work demonstrates that the logarithmic asymptotic
regime starts ``very early'' in the complex plane, already in the region close to the non-perturbed QNM spectrum,
as illustrated in fig.~\ref{fig:log-asymptotics}. The key observational consequence
is that data analysis strategies based on perturbed BH QNM templates constructed on eq. (\ref{e:log-asymps}),
could be successful in extracting perturbed low overtones, therefore probing small BH physics
through the effective parameters introduced out of eq. (1) in the main text.

\bibliographystyle{spmpsci}
\bibliography{Biblio}

\end{document}